\documentclass[%
 reprint,
 amsmath,amssymb,
 aps,
]{revtex4-2}

\usepackage{graphicx}% Include figure files
\usepackage{dcolumn}% Align table columns on decimal point
\usepackage{bm}% bold math
\usepackage{amsmath}
\usepackage{float}
\usepackage{xcolor}
\begin{document}

\preprint{APS/123-QED}

\title{Phase transitions in the Prisoner’s Dilemma game on the Barabási-Albert graph with participation cost}

\author{Jacek Mi\c{e}kisz}\thanks{miekisz@mimuw.edu.pl}
\author{Javad Mohamadichamgavi}\email{jmohamadi@mimuw.edu.pl}
\affiliation{University of Warsaw, Institute of Applied Mathematics and Mechanics, ul. Banacha 2, 
02-097 Warsaw, Poland}

\begin{abstract}
We examine the impact of the maintenance cost of social links on cooperative behavior in the Prisoner’s Dilemma game on the Barabási-Albert scale-free network with a pairwise stochastic imitation. We show by means of Monte Carlo simulations and pair approximation that the cooperation frequency changes abruptly from an almost full cooperation to a much smaller value when we increase the cost of maintaining links. In the critical region, the stationary distribution is bi-modal and the system oscillates between two states: the state with almost full cooperation and one with coexisting strategies. We show that the critical region shrinks with the increasing size of the population. However, the expected time the system spends in a metastable state before switching to the other one does not change as a function of the system's size, which precludes the existence of two stationary states in the thermodynamic limit of the infinite population.
 
\end{abstract}

\maketitle

%\tableofcontents

\emph{Introduction}. One of the fundamental problems of evolutionary biology is understanding the origins of the prevalence of altruistic behavior in the world governed by the Darwinian rule of survival of the fittest \cite{hamilton}. One can discuss it within the framework of evolutionary game theory which describes behavior of systems of many interacting individuals \cite{maynard}. In evolutionary games, players have at their disposal various strategies, receive payoffs which can be interpreted as the number of offspring who inherit their strategies. Evolution of such populations can be modeled by deterministic replicator dynamics or stochastic processes of games on graphs which describe time changes of fractions of populations playing given strategies. 

Famous Prisoner's Dilemma game provides us with an archetypal case of the social dilemma where individual rationality is at odds with the well-being of the population \cite{dresher,axelrod,poundstone}.
In such a two-player game with two strategies, cooperation and defection, a player is tempted to defect, and hence both players defect but they would be better off if they both cooperated. We say that defection is a dominant strategy (it gives the highest payoff regardless of the opponent's strategy). As a result, full defection is the globally asymptotically stable state of the replicator dynamics of the Prisoner's Dilemma game. However, it was shown that in games on graphs (where the payoff of any player is a sum of payoffs resulting from games with its neighbors), cooperative players may form clusters which are resilient against defectors and this leads to the coexistence of both strategies in the stationary state \cite{nowak1,nowak2}. In particular, a scale-free Barabási-Albert random graph \cite{barabasi1,barabasi2}, built by a preferential attachment rule, enhances cooperation. In \cite{pacheco} it was shown that for a certain region of game parameters, stochastic imitation of a better strategy leads to the stationary state of the population with almost all cooperators. An important role of maintaining cooperation is played here by hubs (vertices with high degrees). 

It was shown independently in \cite{sulkowski,masuda} (the expanded version of \cite{sulkowski} was later published in \cite{bpm}) that when we introduce the cost of maintaining a link between neighbors, the level of cooperation in the stationary state decreases. Here we construct a stochastic process (an ergodic Markov chain) on the Barabási-Albert graph where agents follow the Boltzmann updating (called a Fermi rule in the game literature) - they imitate their neighbors with a probability proportional to the exponent of a payoff difference divided by the noise factor, "temperature" of the system. We perform stochastic Monte Carlo simulations and show that the cooperation level abruptly changes from an almost full cooperation to a much smaller value when we increase the cost of maintaining links. We examine closely what happens around a critical value of the cost. The unique stationary probability distribution of our Markov chain is bi-modal in a small interval around the critical cost (bigger the size of the population, smaller the interval) - the population oscillates between peaks of the distribution. 

The situation is analogous to the behavior of the ferromagnetic Ising model on the square lattice at the zero magnetic field and low temperatures, where in the finite volume the system alternates between almost spin-up and spin-down configurations.

However, the situation is different in our case. We inferred from simulations that the expected time the system spends in a metastable state before switching to the other one does not change as a function of the system's size. This precludes the existence of two stationary states in the thermodynamic limit of the infinite population, as is the case in the Ising model.

We construct a pair approximation, a system of differential equations for the frequency of cooperators and the frequency of pairs of cooperators. We solve numerically our equations and show that convergence of solutions changes abruptly at the same critical cost as is present in stochastic simulations.

\emph{Prisoner's Dilemma on the Barabási-Albert graph.} We are concerned here with the two-player Prisoner's Dilemma game with two strategies, cooperation (C) and defection (D), with a canonical parameterization and in addition a cost $\gamma$ of maintaining a link paid by both connected players \cite{sulkowski,masuda,bpm}. Our payoff matrix reads as follows,

\begin{center}
\begin{tabular}{ c c c c }
 &      & C & D \\ 
    & C & $1-\gamma$ & $-\gamma$ \\
$U$ = & & &     \\  
    & D& $T-\gamma$ & $-\gamma$    
\end{tabular}
\end{center}

where the entry $U_{xy}$ is the payoff of the row player using the $x$ strategy while the column player uses the $y$ one.

We set $T > 1$ hence D is a dominant strategy - it gives the highest payoff regardless of the strategy of the opponent - hence a pair $(D,D)$ is the unique Nash equilibrium of the game and the population consisting of just defectors is the globally asymptotically stable state of the replicator dynamics \cite{taylor,hof,weibull}.   

Here we put players on vertices of the Barabási-Albert graph. It is built by the preferential attachment procedure. We start with $m_{o}$ fully connected vertices and then we add $N - m_{o}$ vertices, each time connecting them with $m$ already available vertices with probabilities proportional to their degrees. If $m_{o} = \alpha + 1$ and $m = \alpha /2$, then we get a graph with the average degree equal to $\alpha$. It is known that such a graph in the limit of an infinite number of vertices,
$N$, is scale-free with the probability distribution of degrees given by $p(k) \sim k^{-3}$ \cite{barabasi1,barabasi2}.

Now we present a stochastic dynamics of a spatial game. At discrete moments of time, all individuals interact with their neighbors
and receive payoffs which are sums with respect to individual games. Then the imitation process takes place. A randomly chosen player who has a strategy $x$ and a total payoff $\pi_{x}$ choses randomly one of its neighbors. Assume that a chosen neighbor has a strategy $y$ and a total payoff $\pi_{y}$. The first individual then with the probability $1-\epsilon$ imitates the strategy of the second individual with the probability $w(x \rightarrow y)$, given by the Boltzmann expression:

\begin{equation}
w(x\rightarrow y)=\dfrac{e^{\beta \pi_{y}}}{e^{\beta \pi_{x}}+e^{\beta \pi_{y}}}
\label{up1}
\end{equation}
and with the probability $\epsilon$ adopts the other one. 

The parameter $\beta$ represents a bias toward higher payoffs, better-performing individuals are more likely to be imitated. It also quantifies a noise in the update process, in the limit $\beta \to 0$, individuals are making random updates, and in the limit $\beta \to \infty$, the process becomes deterministic imitation. We set $\epsilon=0.001$,
$\beta=100$, and the average vertex degree, $ \alpha = \langle k \rangle = 4$. For all results, the size of population is $10000$, except where it is varied and explicitly stated.

In this way we constructed a Markov chain with $2^N$ states (assignments of strategies to graph vertices) and the above described transition probabilities. Our Markov chain is irreducible (one can get from any state to any other state in a finite number of steps) and aperiodic which follows from the fact that there is a nonzero probability to stay at any state, and therefore it is ergodic and hence it possesses the unique stationary probability distribution. 

\emph{Phase transition -  Monte Carlo simulations.} We performed stochastic Monte Carlo simulations to estimate the frequency of cooperation in the stationary state. As an initial condition we assign randomly, that is with the probability 1/2, cooperators and defectors to $N$ vertices. To arrive reasonably close to the stationary state, we run simulations for $10^5$ Monte Carlo rounds (each round consists of $N$ steps so each vertex can be on average updated), followed by $10^3$ rounds to compute an average of the cooperation level. We performed $10^3$ runs per every parameter set, averaging over different random initial conditions and network realizations. 

\begin{figure}
    \centering
    \includegraphics[scale=0.55]{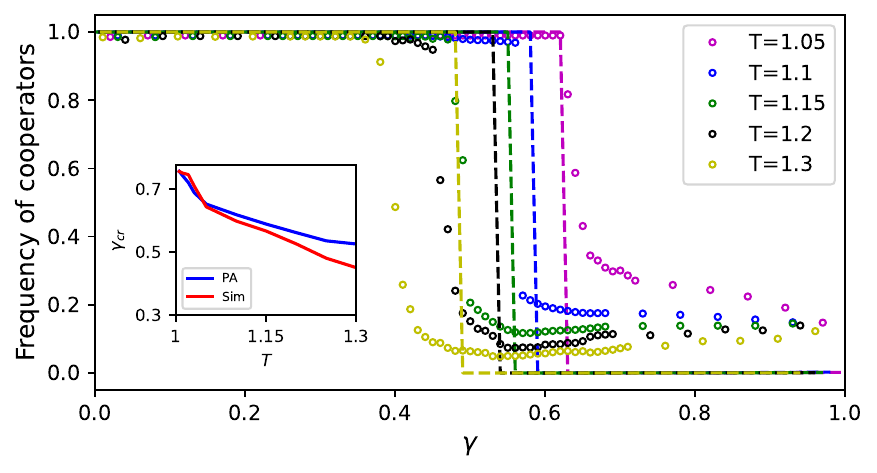}
    \caption{Frequency of cooperators for various values of $T$ as a function of $\gamma$ - the cost of maintaining a link. Dots show the results of simulations and dashed lines indicate stationary states of pair approximation in  eqs. (7-8). In the inset, the critical $\gamma_{cr}$ is shown as a function of $T$.}
    \label{fig:Tand gamma}
\end{figure}

Fig. \ref{fig:Tand gamma} presents the cooperation level in the stationary state as a function of the  link cost for various $T$ values. A sharp transition occurs on an interval $[\gamma_{l}(T,N), \gamma_{r}(T,N)]$ where the system shifts from the full cooperation to the coexistence of cooperation and defection. For low costs we observe a full cooperation which is in agreement with results in \cite{pacheco} for the zero cost. Critical $\gamma$ defined as $\gamma_{cr}(T,N) = (\gamma_{l}(T,N) + \gamma_{r}(T,N))/2$ is a decreasing function of $T$, see the inset in Fig. \ref{fig:Tand gamma} and Fig.
\ref{size}. We see that the critical value of the cost stabilizes as $N$ increases. We conjecture that 
$\lim_{N \rightarrow \infty} \gamma_{l}(T,N) = \lim_{N \rightarrow \infty} \gamma_{l}(T,N) = \gamma_{cr}(T)$. 

This may be compared to the behavior of the ferromagnetic Ising model on the Barabási-Albert graph investigated in \cite{holyst,bianconi}. It was shown there that the effective critical temperature (below which the system is magnetized) diverges as the logarithm of the size of the graph hence the system does not exhibit a phase transition at the thermodynamic limit.

\begin{figure}
    \centering
    \includegraphics[scale=0.45]{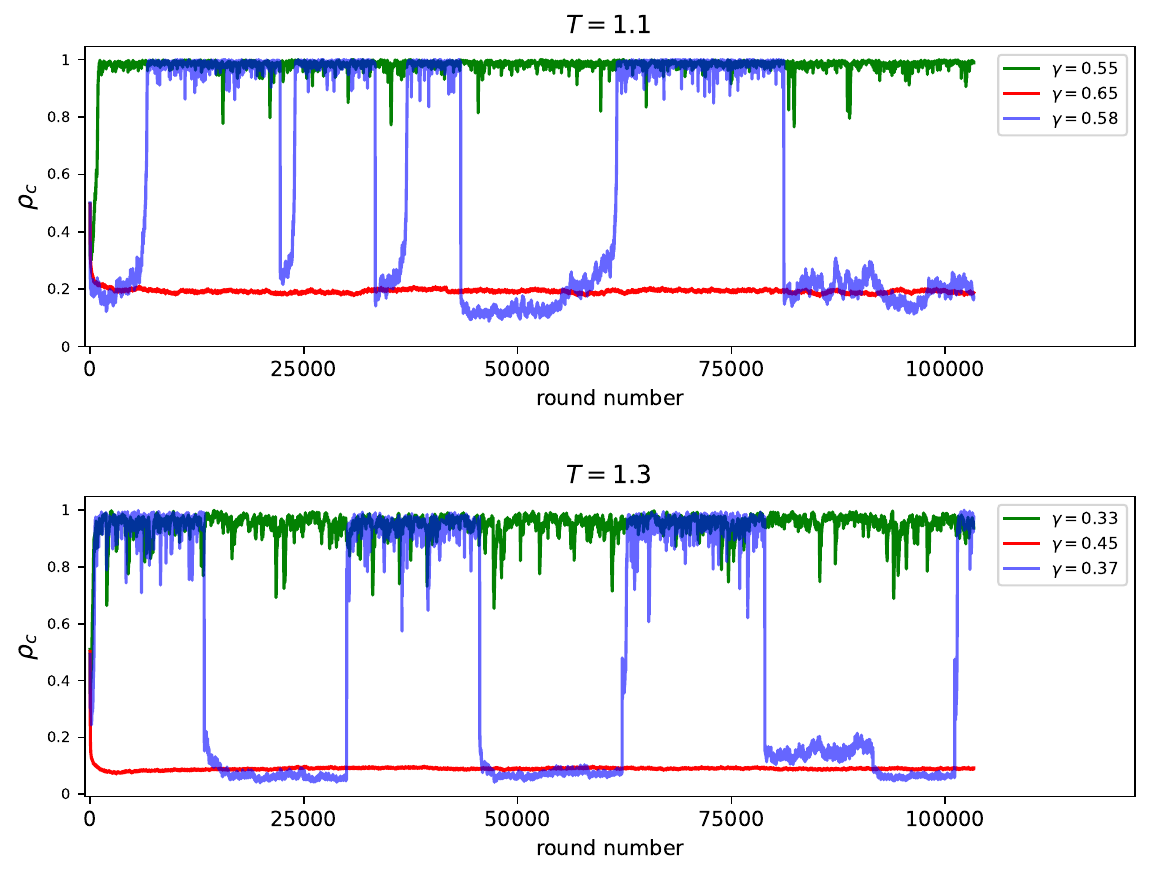}
    \caption{Frequency of cooperators after each round in exemplary simulations for various values of $\gamma$ with $T =1.1$ and $1.3$.}
    \label{fig:timwev2}
\end{figure}

Fig. \ref{fig:timwev2} shows the time evolution of cooperation for $T=1.1$ and $1.3$ for three $\gamma$ values: before, during, and after the phase transition. For $\gamma < \gamma_{l}(T,N)$ cooperation dominates. 
For $\gamma > \gamma_{r}(T,N)$ cooperation and defection coexist, while for $\gamma \in (\gamma_{l}(T,N), \gamma_{r}(T,N))$, the system oscillates between these states. 

\begin{figure}
    \centering
    \includegraphics[scale=0.32]{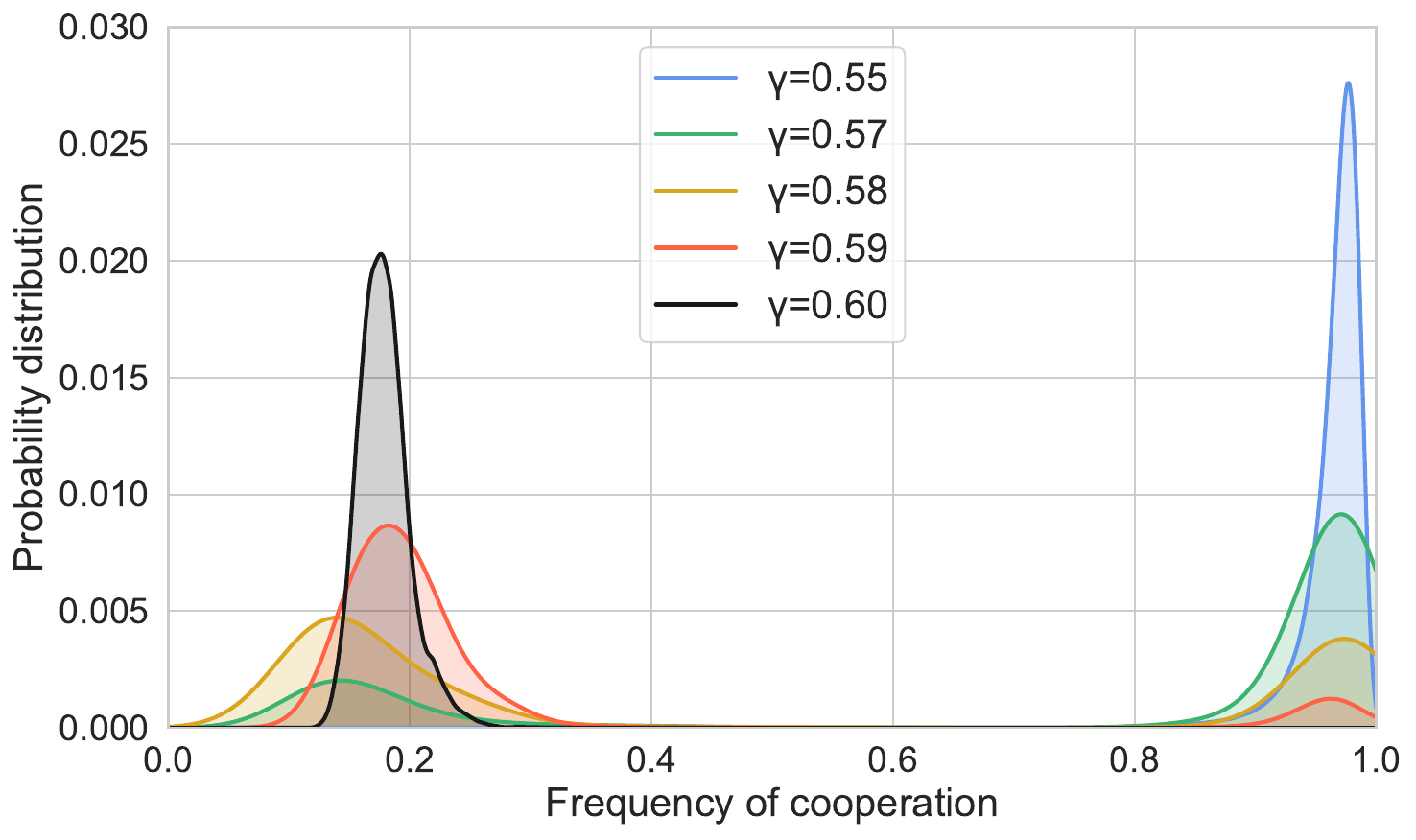}
    \caption{Histogram of the stationary probability distribution of cooperators frequency for various values of $\gamma$ and $T = 1.1$}
    \label{fig:dist}
\end{figure}

\begin{figure}[H]
{\includegraphics[scale=0.65]{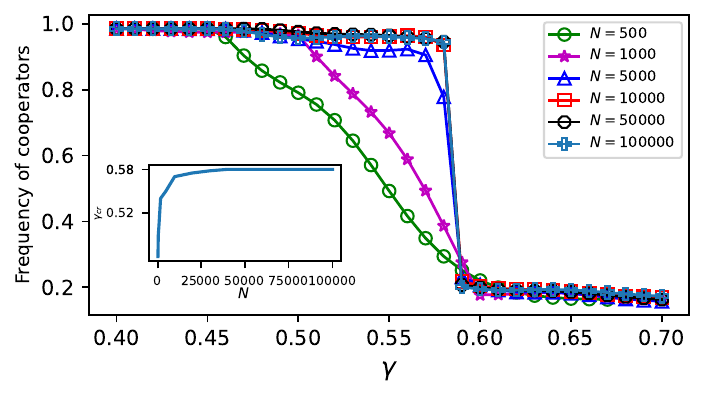}}
\centering
\caption{The effect of $\gamma$  on the frequency of cooperators for various population sizes of the population, $T=1.1.$ In the inset we present the critical cost as a function of N.}
\label{size}
\end{figure}

In Fig. \ref{fig:dist}, we present the stationary distribution of the cooperation frequency for $\gamma \in [\gamma_{l}(T,N),\gamma_{r}(T,N)].$. It is concentrated respectively on full cooperation and coexistence of cooperators and defectors at the segment ends and it is bi-modal in between. This behavior is analogous to that in the two-dimensional ferromagnetic Ising model at the zero magnetic field, where in a finite volume we observe switching between two ensembles of most up and most down spins, that is almost $1$ and almost $-1$ magnetization per lattice site.

These oscillations and the previously discussed stabilization of the critical cost with respect to the size of the system may suggest that in the thermodynamic limit we could have two stationary states like in the Ising model. However, it is not the case here. We inferred from simulations that the expected time the system spends in a metastable state before switching to the other one does not change as a function of the system's size. This precludes the existence of two stationary states. It is caused by the existence of hubs (vertices with a very high degree) in our model. 

Notably, for low $\gamma$ cooperation initially declines. If a cooperator in a highly connected hub becomes a defector, its high payoff accelerates its strategy spread, triggering a temporary cooperation decline. As defection expands, the payoff advantage fades, enabling cooperators to reclaim dominance. For $\gamma$ in the above described interval, cooperation exhibits metastability, fluctuating between high and low cooperation states. Hubs play a crucial role, if a hub remains cooperative, cooperation thrives, but with a probability $\epsilon$, it may switch to defection, tipping the balance. This leads to  metastable states, where transitions occur as hubs alter their strategies.

\emph{Phase transition - pair approximation.} In pair approximations, a state of a population of players on graphs is represented by frequencies of strategies and pairs of nearest-neighbor strategies.
We denote by $\rho_C$ and $\rho_D$ a frequency of cooperators (C) and defectors (D), respectively and by $\rho_{CC}$, $\rho_{CD}$, and $\rho_{DD}$, frequencies of pairs of players. These variables satisfy the following relations: $\rho_C + \rho_D =1$,
$\rho_{CC} + \rho_{CD} + \rho_{DD} = 1$, and $\rho_{C} = \rho_{CC} + (\rho_{CD})/2$. Hence the system can be described by just two variables, say $\rho_{C}$ and $\rho_{CC}$. Below we derive differential equations for these variables.

Consider a randomly chosen individual with strategy $x$, having $k$ neighbors. It is connected to $k_{x}$ individuals with the same strategy with probability $P_{x \vert x}$, and to $k-k_{x}$ individuals with the opposite strategy with probability $1-P_{x \vert x}$. The probability of such local configuration is described by a binomial distribution,
\begin{equation}
\Phi(k,k_{x})=\binom{k}{k_{x}} {P_{x\vert x}}^{k_{x}} (1-P_{x\vert x})^{k-k_{x}}
\label{bio1}
\end{equation}
In each imitation step, we first select with probability $\rho_x$ an individual with a strategy $x$. Next, we choose for imitation a connected individual with the other strategy $y$ with a probability $\dfrac{k-k_{x}}{k}$, where $k$ is the number of connections of the first individual. The probability of this specific configuration is given by $\Phi(k,k_{x})$. 

The payoff of the first focal individual with strategy $x$ is equal to 
\begin{equation}
    \pi_{x}(k,k_{x})=k_{x}U_{xx}+(k-k_{x})U_{xy},
\end{equation} 
where we have taken into account that it is connected to $k_{x}$ individual with the same strategy with probability given in \eqref{bio1}.

To compute the payoff of the second individual near the focal individual, denoted as $\pi_{y}^{\prime}$, we account for the fact that in homogeneous networks without degree-degree correlation, the degree distribution of a randomly chosen neighbor follows $kp(k)$ rather than $p(k)$ \cite{friends}. When an individual with degree $k$ adopts strategy $y$ and is a neighbor of a focal individual with strategy $x$, they must have at least one $x$ neighbor, while the remaining $k-1$ neighbors follow the distribution $\Phi(k-1,k_{y})$. Thus the average payoff of a neighbor of the focal individual is given by the following expressions for $y \in \{D, C\}$:

\begin{align}
\pi_C^{{\prime}} &= \sum\limits_{k=1}^{N} \dfrac{k p(k)}{\langle k\rangle} \Bigg(\sum\limits_{k_C=0}^{k\!-\!1}\Phi(k\!-\!1,k_C) \Big(k_C(1\!-\!\gamma)\nonumber \\
&\!-\!(k\!-\!1\!-\!k_C)\gamma\Big)\!-\!\gamma\Bigg)= \frac{\langle k^2 \rangle}{\langle k \rangle}(P_{C\vert C}\!-\!\gamma)\!-\!P_{C\vert C},\nonumber \\
\pi_D^{{\prime}} &=\sum\limits_{k=1}^{N}\dfrac{k p(k)}{\langle k\rangle}  \Bigg(\sum\limits_{k_D=0}^{k\!-\!1}\Phi(k\!-\!1,k_D) \Big((k\!-\!1\!-\!k_D)(T\!-\!\gamma)\nonumber\\
&\!-\!k_D\gamma\Big)+(T\!-\!\gamma)\Bigg) = \frac{\langle k^2 \rangle}{\langle k \rangle}(T(1\!-\!P_{D\vert D})\!-\!\gamma)+TP_{D\vert D}~.
\end{align}

where $<k^2>$ is the second moment of degree and define as $<k^2>=\sum\limits_{k=1}^{N} k^2 p(k)$. 

The first individual, with a payoff of $\pi_{x}(k,k_{x})$, then imitates the strategy of its neighbor, who has an average payoff of $\pi_{y}^{'}$, based on the probability $w$ given in equation (\ref{up1}), denoted as $P_{y \rightarrow x}(k,k_{x})$. For $x, y \in \{C, D\}$, we have:

\begin{subequations}\label{PA}
    \begin{align}
        P_{D \rightarrow C}(k,k_D) &=  
     \frac{(1-\epsilon) \rho_D \frac{k-k_D}{k} \Phi(k,k_D)}{1+e^{\beta(\pi_D(k,k_D)-\pi_C^{'})}} + \epsilon \rho_D
    \label{PA1}\\
        P_{C \rightarrow D}(k,k_C) &= 
     \frac{(1-\epsilon)\rho_C \frac{k-k_C}{k} \Phi(k,k_C) }{1+e^{\beta(\pi_C(k,k_C)-\pi_D^{'})}}+ \epsilon \rho_C.
    \label{PA2}
    \end{align}
\end{subequations}

To determine the frequency of $CC$ pairs over time, consider a cooperator with \( k \) neighbors, \( k_C \) of whom are cooperators. If replaced by a defector (probability \( P_{C \rightarrow D} \)), CC pairs decrease by \( k_C \). If a defector with \( k_D \) defector neighbors switches to cooperation (\( P_{D \rightarrow C} \)), CC pairs increase by \( k - k_D \). These changes occur with the following rates:
\begin{subequations}\label{PAA}
    \begin{align}
        P_{CD \rightarrow CC}(k,k_D)&=P_{D \rightarrow C}(k,k_D)(k-k_D)
    \label{PAA1},\\
        P_{CC \rightarrow CD}(k,k_C)&=P_{C \rightarrow D}(k,k_C)(k_C).
    \label{PAA2}
    \end{align}
\end{subequations}
 
Now we can write differential equations for the frequency of cooperators $\rho_{C}$ and pairs of cooperators $\rho_{CC}$,
\begin{align}
     \frac{d\rho_C}{dt}&=\sum\limits_{k=1}^{\infty}p(k)\left(\sum\limits_{k_D=0}^{k}P_{D \rightarrow C}(k,k_D)-\sum\limits_{k_C=0}^{k}P_{C \rightarrow D}(k,k_C)\right)\label{pc1}
\end{align}

\begin{align}
     \frac{d\rho_{CC}}{dt}&=\sum\limits_{k=1}^{\infty}p(k)\Bigg(\sum\limits_{k_D=0}^{k}P_{CD \rightarrow CC}(k,k_D)\nonumber\\&-\sum\limits_{k_C=0}^{k}P_{CC \rightarrow CD}(k,k_C)\Bigg)\label{pcc1}.
\end{align}

We would like to see how good pair approximation is with respect to Monte Carlo simulations. To compare it with the behavior of a system with $N$ players
we constrain summations in \eqref{pc1} and \eqref{pcc1} by the maximal degree and we use $p(k)$ obtained in simulations. We set initial condition as $\rho_{C}=0.5$ and $\rho_{CC}=0.25$ (based on random distribution of strategies) and solve the system of differential equations numerically (we used the fourth-order Runge-Kutta method) to find approximate stationary states, when both $\rho_{C}$ and $\rho_{CC}$ almost do not change.

Results are presented in Fig.1. We observe a sharp phase transition in the cooperation level

We define the critical value of $\gamma$ predicted by pair approximation as the average of the two $\gamma$ values — one corresponding to almost full cooperation and the other to almost total defection — measured just before and after the phase transition. For  example, when $T = 1.1$, the system exhibits full cooperation at $\gamma = 0.5901$, but a slight increase to $\gamma = 0.5902$ leads to almost total defection. Therefore, we define the critical value as $\gamma_{cr} = 0.59015$, which is in good agreement with the simulation $\gamma_{cr} = 0.58$.   
 
 We see that our pair approximation works better for small $T$ which is consistent with previous findings. Additionally, the pair approximation performs better for smaller $\gamma$'s. 

\emph{Discussion.} We analyzed the stochastic imitation dynamics of the Prisoner's Dilemma game on the Barabási-Albert graph with a participation cost introduced in \cite{sulkowski,masuda,bpm}. We showed by means of Monte Carlo simulations the existence of a critical cost region with an abrupt decrease of the cooperation level in the stationary state. In this region, the stationary distribution is bi-modal and the system oscillates between two states: the state with almost full cooperation and the state with coexisting strategies. We show that the critical region shrinks with the increasing size of the population.

We designed a pair approximation - the system of two differential equations for the frequency of cooperators and nearest-neighbor pair of cooperators. The stationary state of such a system shows an abrupt change from full cooperation to full defection.  

Such a behavior is reminiscent of the discontinuous first-order phase transition in the two-dimensional ferromagnetic Ising model on the square lattice. 
However, we inferred from simulations that the expected time the system spends in a metastable state before switching to the other one does not change as a function of the system's size, which precludes the existence of two stationary states, that is, breaking of ergodicity. This is caused by the existence of hubs in the Barabási-Albert scale-free graphs. Nevertheless, we observe a sharp phase transition for any finite size of the graph. This seems to be a novel behavior in systems of many interacting entities.

It is worth mentioning here that the ferromagnetic Ising model on the Barabási-Albert graph was discussed in \cite{holyst,bianconi}. It was shown both by stochastic simulations and the mean-field approximation that the effective critical temperature diverges as the logarithm of the size of the population and hence in the infinite-population thermodynamic limit the system does not exhibit a phase transition. We would like to point out that this model satisfies the detailed balance, so it is time reversible. In contrast, our imitation dynamics is time irreversible. 

It is a fundamental problem to investigate what are necessary and sufficient conditions of the underlying graph and dynamics to imply the existence of phase transitions.

{\bf Acknowledgments}: This project has received funding from the European Union Horizon 2020 research and innovation program
under the Marie Sk\l odowska-Curie grant agreement No 955708. Computer simulations were made with the support of the Interdisciplinary Center for Mathematical and Computational Modeling of the University of Warsaw (ICM UW).

\end{document}